\pdfoutput=1
\documentclass[aps,prl,superscriptaddress,twocolumn,twoside]{revtex4}
\usepackage[utf8]{inputenc}

\usepackage{mathrsfs,mathtools,bbold}
\usepackage{amsmath,amssymb}
\usepackage{amsfonts}
\usepackage{verbatim}
\usepackage{graphicx}
\usepackage{mathrsfs,mathtools}
\usepackage{ccaption}
\usepackage[dvipsnames]{xcolor}
\definecolor{SchoolColor}{rgb}{0.6471, 0.1098, 0.1882} 
\usepackage{stmaryrd}
\usepackage{textcomp}
\usepackage{eucal} 
\usepackage{setspace}
\usepackage{empheq}
\usepackage[colorlinks=true,linkcolor=black,citecolor=teal,urlcolor=MidnightBlue,filecolor=black]{hyperref}

\DeclareFontFamily{OT1}{rsfs}{}
\DeclareFontShape{OT1}{rsfs}{m}{n}{ <-7> rsfs5 <7-10> rsfs7 <10->rsfs10}{} 
\DeclareMathAlphabet{\mycal}{OT1}{rsfs}{m}{n}

\newcommand{\unity}{1\hspace{-0.243em}\text{l}}

\newcommand{\be}[1]{ \begin{equation}\label{#1} }
\newcommand{\ee}{\end{equation}}
\newcommand{\bea}[1]{\begin{eqnarray}\label{#1} }
\newcommand{\eea}{\end{eqnarray}}

\newcommand{\tr}{\textrm{tr}}

\newcommand{\eq}[2]{\begin{equation} #1 \label{#2} \end{equation}}
\newcommand{\eps}{\varepsilon}

\DeclareMathOperator{\extdm}{d}
\newcommand{\extd}{\extdm \!}

\begin{document}

\title{Flat space holography and complex SYK}

\author{Hamid Afshar}
\email[]{afshar@hep.itp.tuwien.ac.at}

\affiliation{Institute for Theoretical Physics, TU Wien, Wiedner Hauptstr. 8, A-1040 Vienna, Austria}
\affiliation{School of Physics, Institute for Research in Fundamental Sciences (IPM),
P.O.Box 19395-5531, Tehran, Iran}

\author{Hern\'an A. Gonz\'alez}
\email[]{hernan.gonzalez@uai.cl}

\affiliation{Facultad de Artes Liberales, Universidad Adolfo Ib\'a\~nez, Diagonal Las Torres 2640, Pe\~nalol\'en, Santiago, Chile}

\author{Daniel Grumiller}
\email[]{grumil@hep.itp.tuwien.ac.at}

\affiliation{Institute for Theoretical Physics, TU Wien, Wiedner Hauptstr. 8, A-1040 Vienna, Austria}

\author{Dmitri Vassilevich}
\email[]{dvassil@gmail.com}

\affiliation{CMCC, Universidade Federal do ABC, Santo Andre, Sao Paulo, Brazil}
\affiliation{Physics Department, Tomsk State University, Tomsk, Russia}

\date{\today}

\begin{abstract}
We provide the first steps towards a flat space holographic correspondence in two bulk spacetime dimensions. The gravity side is described by a conformally transformed version of the matterless Callan--Giddings--Harvey--Strominger model. The field theory side follows from the complex Sachdev--Ye--Kitaev model in the limit of large specific heat and vanishing compressibility. We derive the boundary action analogous to the Schwarzian as the key link between gravity and field theory sides and show that it coincides with a geometric action discovered recently by one of us \cite{Afshar:2019tvp}. 
\end{abstract}

\maketitle

\hypersetup{linkcolor=SchoolColor}

\newcommand{\lo}{{\cal P}}
\newcommand{\nlo}{{\cal T}}
\newcommand{\zen}{{\cal Z}}

\paragraph{Introduction.} The Sachdev--Ye--Kitaev (SYK) model \cite{Kitaev:15ur,Sachdev:1992fk,Sachdev:2010um} reinvigorated studies of Jackiw--Teitelboim (JT) gravity \cite{Jackiw:1984,Teitelboim:1984} since in a certain limit it is the gravity dual of the former \cite{Maldacena:2016hyu}. This holographic relationship, dubbed nAdS$_2$/nCFT$_1$, inspired numerous research activities in the past few years in gravity and condensed-matter communities, see e.g.~\cite{Grumiller:2015vaa,Polchinski:2016xgd,You:2016ldz,Jevicki:2016bwu,Jensen:2016pah,Maldacena:2016upp,Engelsoy:2016xyb,Bagrets:2016cdf,Garcia-Alvarez:2016wem,Cvetic:2016eiv,Jevicki:2016ito,Gu:2016oyy,Gross:2016kjj,Berkooz:2016cvq,Garcia-Garcia:2016mno,Banerjee:2016ncu,Fu:2016vas,Witten:2016iux,Cotler:2016fpe,Klebanov:2016xxf,Davison:2016ngz,Peng:2016mxj,Krishnan:2016bvg,Turiaci:2017zwd,Bi:2017yvx,Li:2017hdt,Gurau:2017xhf,Mandal:2017thl,Gross:2017hcz,Caputa:2017urj,Jian:2017unn,Song:2017pfw,Mertens:2017mtv,Krishnan:2017txw,Garcia-Garcia:2017bkg,Azeyanagi:2017drg,Grumiller:2017qao,Haehl:2017pak,Altland:2019lne,Hirano:2019iwt}. One crucial piece of the puzzle is the Schwarzian action \cite{Kitaev:15ur,Maldacena:2016hyu} that arises in the large $N$ and strong coupling limit on the quantum mechanics side and, upon imposing suitable boundary conditions, also on the gravity side.

Given the impressive evidence for AdS/CFT-realizations of holography it is justified to ask how general the holographic principle works, if it works beyond the AdS/CFT correspondence and in particular if and how it works in asymptotically flat spacetimes. See \cite{Polchinski:1999ry,Susskind:1998vk,Giddings:1999jq,Mann:2005yr,Dappiaggi:2005ci,Barnich:2009se,Bagchi:2010eg,Bagchi:2012yk,Barnich:2012xq,Bagchi:2012xr,Barnich:2013axa,Bagchi:2014iea,Bagchi:2016bcd,Bagchi:2016geg,Jiang:2017ecm,Ciambelli:2018wre,Hijano:2018nhq,Fareghbal:2018ngr,Ball:2019atb,Hijano:2019qmi,Grumiller:2019xna,Godet:2019wje,Fareghbal:2019czx,Banks:2015mya} and refs.~therein for selected earlier results in flat space holography. 

The main goal of our Letter is to find a model analogous to JT that leads to a holographic relationship involving flat space instead of AdS$_2$, with some suitable replacement of the Schwarzian action.

The Callan--Giddings--Harvey--Strominger (CGHS) model \cite{Callan:1992rs} is a prime candidate for the gravity side of flat space holography since all solutions have asymptotically vanishing Ricci scalar. This opens up the prospect to construct a concrete holographic correspondence between flat space dilaton gravity in 1+1 dimensions and some cleverly designed quantum system of (complex) fermions in 0+1 dimensions. 

The principal result of this Letter is that the flat space analogue of the Schwarzian action is given by
\eq{
I^{\textrm{\tiny{tw}}}[h,\,g] = \kappa\,\int\limits_0^\beta{\extd\tau}\,\bigg(\nlo\,h^{\prime\,2}  -  g'\bigg(i\lo\,h^\prime + \frac{h''}{ h'}\bigg) + g''
\bigg)
}{eq:intro}
where $\kappa$ is a coupling constant, $\beta$ is inverse temperature, and prime denotes derivative with respect to $\tau$, the time direction along the boundary. The time-reparametrization field $h(\tau+\beta)=h(\tau)+\beta$ is quasi-periodic and the phase field $g(\tau)$, in the absence of winding, is periodic. When the functions $\nlo$ and $\lo$ are constant we refer to them as mass and charge, respectively. While mass can be arbitrary it will turn out that regularity demands a linear relationship between charge and temperature. The superscript {\tiny{tw}} stands for `twisted warped' and stems from the symmetries \eqref{eq:cghs40too} that govern our action. On the gravity side $\kappa$ is essentially the inverse Newton constant, as evident from our starting point \eqref{eq:intro3}. On the field theory side $\kappa$ is essentially the geometric mean of specific heat at constant charge and zero temperature compressibility, as evident from our final equation \eqref{eq:cghs102}.

The remainder of our Letter is organized as follows. We start by gaining some intuition about our gravity model in the metric formulation and then switch to a gauge theoretic formulation. The latter is employed to derive our main result \eqref{eq:intro}. Finally, we recover the boundary action \eqref{eq:intro} from a scaling limit of complex SYK.

\newcommand{\cghs}{\widehat{\textrm{\tiny CGHS}}}

\paragraph{Metric formulation of CGHS.} Following Cangemi and Jackiw \cite{Cangemi:1992bj} we manipulate the CGHS action \cite{Callan:1992rs} in three ways: 1.~for simplicity we set all matter fields to zero, 2.~we perform a Weyl rescaling (depending on the dilaton $X$) of the metric $g_{\mu\nu}$, and 3.~we ``integrate in'' an abelian gauge field $A_\mu$ and an auxiliary scalar field $Y$ that is constant on-shell. The action ($\varepsilon^{\mu\nu}$ is the $\varepsilon$-tensor)
\eq{
I_{\cghs} = \frac{\kappa}{2}\,\int\extd^2x\sqrt{-g}\,\big(XR-2Y + 2Y \varepsilon^{\mu\nu}\partial_\mu A_\nu\big)
}{eq:intro3} 
provides a reformulation of the CGHS model referred to as $\cghs$. We solve now the $\cghs$ field equations
\begin{align}
    R &= 0 \label{eq:cghs42} \\
    \varepsilon^{\mu\nu}\partial_\mu A_\nu &= 1 \label{eq:cghs73} \\
    \nabla_\mu\nabla_\nu X - g_{\mu\nu}\nabla^2 X &= g_{\mu\nu} Y \qquad Y=\Lambda = \rm const.\label{eq:cghs72} 
\end{align}
with suitable boundary and gauge fixing conditions.

In Eddington--Finkelstein gauge the most general solution to the Ricci-flatness condition \eqref{eq:cghs42} is given by Rindler-type black hole metrics of the form
\eq{
\extd s^2 = -2\extd u \extd r + 2\big(\lo(u)\,r + \nlo(u)\big)\,\extd u^2\,.
}{eq:cghs41}

The loosest set of boundary conditions compatible with the gauge fixing \eqref{eq:cghs41} allows fluctuations of both free functions, $\delta\lo\neq0\neq\delta\nlo$. These boundary- and gauge-fixing conditions are preserved by asymptotic Killing vectors $\xi(\varepsilon,\,\eta)=\varepsilon(u)\,\partial_u-\big(\varepsilon'(u)r+\eta(u)\big)\,\partial_r$ since ${\cal L}_\xi g_{\mu\nu}={\cal O}(\delta g_{\mu\nu})$, namely ${\cal L}_\xi g_{rr}=0={\cal L}_\xi g_{ur}$ and ${\cal L}_\xi g_{uu}=\delta_\xi\lo\,r + \delta_\xi\nlo$, with $\delta_\xi\lo = \varepsilon\lo' + \varepsilon'\lo + \varepsilon''$ and $\delta_\xi\nlo = \varepsilon\nlo' + 2\varepsilon'\nlo  + \eta' - \eta\lo$. Prime is the derivative along retarded time $u$ 
and ${\cal L}_\xi$ the Lie-derivative along $\xi$. 

The Lie-bracket algebra of the asymptotic Killing vectors $[\xi(\varepsilon_1,\eta_1),\,\xi(\varepsilon_2,\eta_2)]_{\textrm{\tiny{Lie}}}=\xi(\varepsilon_1\varepsilon_2'-\varepsilon_2\varepsilon_1',\,(\varepsilon_1\eta_2-\varepsilon_2\eta_1)')$ in terms of Laurent modes, $L_n = \xi(\varepsilon=-u^{n+1},\eta=0)$ and $M_n = \xi(\varepsilon=0,\eta=u^{n-1})$, yields $[L_n,\,L_m]_{\textrm{\tiny Lie}}=(n-m)\,L_{n+m}$, $[L_n,\,M_m]_{\textrm{\tiny Lie}}=-(n+m)\,M_{n+m}$ and $[M_n,\,M_m]_{\textrm{\tiny Lie}}=0$. This algebra consists of a Witt subalgebra generated by $L_n$ and spin-0 super\-trans\-lations generated by $M_n$. We refer to it as BMS$_2$~\footnote{%
The acronym BMS derives from seminal work by Bondi, van der Burgh, Metzner and Sachs \cite{Bondi:1962,Sachs:1962} on asymptotic symmetries of asymptotically flat spacetimes in four spacetime dimensions. In distinction to BMS$_4$ the BMS$_2$ supertranslations are radial. In Euclidean signature with Euclidean time $t\sim{t}+\beta$ instead of Laurent modes Fourier modes are used, $L_n=\xi(\varepsilon=ie^{2\pi int/\beta},0)$ and $M_n=\xi(0,\eta=e^{2\pi int/\beta})$, with no changes in the algebra.}.

In axial gauge for the $U(1)$ connection the field equation \eqref{eq:cghs73} is solved by the two-dimensional Coulomb connection $A = r\,\extd u$. Its preservation under combined diffeomorphisms and gauge transformations, $\delta_{\xi,\sigma} A_\nu = \xi^\mu\partial_\mu A_\nu + A_\mu\partial_\nu\xi^\mu + \partial_\nu\sigma$, relates the functions $\eta$ and $\sigma$, $\eta = \sigma^\prime$, which can be interpreted in two different ways. Either one concludes that $\sigma$ has to contain a $\ln{u}$-term, since $M_0$ is allowed to be non-zero, or one concludes that $M_0$ is forbidden, since $\sigma$ is assumed to have a Laurent series around $u=0$. The first option leads to the BMS$_2$ symmetries discussed above. The second option leads to a slight modification of the transformation properties.
\begin{subequations}
\label{eq:cghs40too}
\begin{align}
  \delta_\xi\lo &= \varepsilon\lo' + \varepsilon'\lo + \varepsilon'' \\
  \delta_\xi\nlo &= \varepsilon\nlo' + 2\varepsilon'\nlo  + \sigma^{\prime\prime} - \sigma^\prime\lo \label{eq:cghs40}
\end{align}
\end{subequations}
In the present Letter we focus on the second option, since it guarantees that the Wilson loop in the complex $u$-plane encircling the origin $u=0$ is gauge invariant, $\delta_\sigma\oint{A}=0$; in other words, there are no winding modes.

Defining instead of $M_n$ new Fourier modes $J_n:=\xi(0,\,\sigma=u^n)$ yields the asymptotic symmetry algebra in terms of asymptotic Killing vector modes 
 $[L_n,\,L_m]_{\textrm{\tiny Lie}} = (n-m)\,L_{n+m}$,
 $[L_n,\,J_m]_{\textrm{\tiny Lie}} = -m\,J_{n+m}$ and
 $[J_n,\,J_m]_{\textrm{\tiny Lie}} = 0$,
which is known as ``warped Witt algebra'', the centerless version of either the warped conformal algebra \cite{Detournay:2012pc} or the twisted warped conformal algebra \cite{Afshar:2015wjm}. 

Finally, the $rr$-component of the field equation \eqref{eq:cghs72} is solved by dilaton fields linear in the radial coordinate
\eq{
X = x_1(u)\,r + x_0(u)
}{eq:cghs65}
The remaining components of the field equation \eqref{eq:cghs72}, which involve the functions $x_i(u)$, will be determined in the gauge theoretic formulation of $\cghs$. 

\paragraph{Gauge theory formulation of CGHS.} For the gauge theoretic formulation as non-abelian BF-theory \cite{Cangemi:1992bj,Verlinde:1991rf} we use conventions analogous to \cite{Gonzalez:2018enk}. The first order form of the $\cghs$ bulk action \eqref{eq:intro3} is given by
\eq{
I^{\textrm{\tiny{BF}}}[B,\,{\cal A}] = \kappa\,\int\langle B,\, F\rangle
}{eq:angelinajolie}
where $\kappa$ is the coupling constant, $B$ is a scalar and $F=\extd{\cal{A}} + {\cal{A}}\wedge{\cal{A}}$ the non-abelian field strength. The connection
\eq{
 {\cal{A}} = \omega\,J + e^a\,P_a + A\,Z
}{eq:cghs1}
contains dualized spin-connection $\omega$, zweibein $e^a$ and $U(1)$ connection $A$.  The generators obey the Maxwell algebra~\footnote{See the Supplemental Material for more on the Maxwell algebra in 1+1 dimensions and e.g.~\cite{Bonanos:2009wy,Matulich:2019cdo} as well as refs.~therein for more on the Maxwell (super-)algebra in higher dimensions.} whose non-zero commutators read $[P_a,\,P_b] = \epsilon_{ab}\,Z$ and $[P_a,\,J] = \epsilon_a{}^b\,P_b$. The scalar field 
\eq{
B = X\,Z + X^a \epsilon_a{}^b\,P_b + Y\, J
}{eq:cghs2}
comprises the dilaton $X$, Lagrange multipliers $X^a$ for torsion constraints and the auxiliary field $Y$. Finally, $\langle\,,\,\rangle$ denotes the bilinear form with non-vanishing entries
\eq{
\langle J,\,Z\rangle = -1 \qquad\qquad \langle P_a,\,P_b\rangle = \eta_{ab}\,.
}{eq:cghs3}
We use light-cone gauge for the Minkowski metric, $\eta_{+-}=1$, in terms of which the Levi-Civit\'a symbol is $\epsilon_\pm{}^\pm=\pm 1$ and the gauge algebra reads $[P_+,\,P_-]=Z$ and $[P_\pm,\,J]={\pm}P_\pm$. 
Integrating out the Lagrange multipliers $X^a$ and solving the torsion constraints, the action \eqref{eq:angelinajolie} with \eqref{eq:cghs1}-\eqref{eq:cghs3} can be shown to be equivalent to \eqref{eq:intro3}.

Boundary conditions compatible with the ones in the metric formulation are given by
\eq{
{\cal A} = b^{-1}\,\big(\extd + a\big)\,b\qquad\qquad B = b^{-1}\,x\,b
}{eq:cghs4}
with $b=\exp{\big(-r\,P_+\big)}$ and
\begin{align}
 a &= \big(\nlo(u)\,P_+ + P_- + \lo(u)\,J \big)\,\extd u \label{eq:ansatz1}\\
 x &= x^+(u)\,P_+ + x_1(u)\,P_- + Y\,J + x_0(u)\, Z \label{eq:ansatz2}
\end{align}
where both functions in the connection are allowed to vary, $\delta\nlo\neq0\neq\delta\lo$.

\newcommand{\feta}{x_0} 
\newcommand{\feps}{x_1} 

The equations of motion reduce to 
\eq{
 \extd a + a\wedge a = 0 =
 \extd x + [a,\,x] \,.
}{eq:eom}
The first one is obeyed automatically by our ansatz \eqref{eq:ansatz1}. The second one, which states that $x$ is the stabilizer of $a$, holds provided $Y = \Lambda = \rm const.$ and the following differential equations are fulfilled: $(x^+)^\prime = \lo\,x^+ - \nlo\,Y$, $x_1^\prime = -\lo\,x_1 + Y$ and $x_0^\prime = x^+ - \nlo\,x_1$. Using them, $x$ is conveniently parametrized by two functions $\feps$ and $\feta$.
\eq{
x = \big(\feta^\prime + \nlo\,\feps\big)\,P_+ + \feps\, P_- + \big(\feps^\prime + \lo \,\feps \big)\,J + \feta\, Z
}{eq:cghs27}

\paragraph{Asymptotic symmetries.} The boundary condition preserving gauge transformations $\delta_\lambda{B}=[B,\lambda]$ and 
\eq{
\delta_\lambda {\cal A} = \extd \lambda + [{\cal A},\lambda] \stackrel{!}{=} {\cal O}(\delta {\cal A}) = \big({\cal O}(r) P_+ + {\cal O}(1) J\big)\extd u
}{eq:cghs17}
are generated by gauge parameters $\lambda=b^{-1}\,\epsilon\,b$ with
\eq{
\epsilon=\epsilon^+(u)\,P_+ + \eps(u)\,P_- + \epsilon^J(u)\,J + \sigma(u)\,Z\,.
}{eq:cghs18}
The absence of the $P_-$- and $Z$-components on the right hand side of \eqref{eq:cghs17} yields consistency relations between the functions in the gauge parameter \eqref{eq:cghs18}.
\eq{
\epsilon^J = \eps^\prime + \lo\,\eps\qquad\qquad \epsilon^+ = \sigma^\prime + \nlo\,\eps
}{eq:cghs20}

The boundary condition preserving gauge transformations \eqref{eq:cghs17}-\eqref{eq:cghs20} imply precisely the transformation laws \eqref{eq:cghs40too}, which is the twisted warped conformal transformation behavior introduced in \cite{Afshar:2015wjm}. Therefore, the analogue of the conformal symmetries that govern the Schwarzian action are twisted warped conformal symmetries, which govern our boundary action \eqref{eq:intro}.

\paragraph{Boundary action.} Our derivation of the boundary action for $\cghs$ follows closely the derivation of the Schwarzian action in section 3 of \cite{Gonzalez:2018enk}. 

From now on we work in Euclidean signature with periodic boundary time, $t\sim{t}+\beta$, where $\beta$ is inverse temperature. Mapping Lorentzian to Euclidean results requires the following replacements: $u\to{iu}=t$, $\eta_{ab}\to\delta_{ab}$, $P_\pm=P^{\textrm{\tiny{E}}}_1\pm{i}P^{\textrm{\tiny{E}}}_0$,  $Z=iZ^{\textrm{\tiny{E}}}$,  $J=iJ^{\textrm{\tiny{E}}}$ and $x_{0,1}=-ix_{0,1}^{\textrm{\tiny{E}}}$ (with real $x_{0,1}^{\textrm{\tiny{E}}}$). The fields are given by \eqref{eq:cghs1} and \eqref{eq:cghs2}, with all quantities replaced by their Euclidean counterparts. We use the definition $\oint\extd{t}=\int_0^\beta\extd{t}$ and dot means $\frac{\extd}{\extd{t}}$.

The variation of the BF-action \eqref{eq:angelinajolie} apart from the bulk equations of motion yields a boundary term \cite{Gonzalez:2018enk}.
\eq{
\delta I^{\textrm{\tiny{tw}}} = -\kappa\,\oint\extd t \,\langle x,\, \delta a_t\rangle 
}{eq:cghs28}
Our aim is to cancel this boundary term by variation of a boundary action. To this end we use a convenient representation of the connection, $a_t=f_tx+G^{-1}\partial_tG$. A consistent choice is $G=\exp{({\feta^{\textrm{\tiny{E}}}}(iP^{\textrm{\tiny{E}}}_1-P^{\textrm{\tiny{E}}}_0))}$ $\exp{(-i\ln(-i{\feps^{\textrm{\tiny{E}}}})J^{\textrm{\tiny{E}}})}$ $\exp(-\int^t\feta^{\textrm{\tiny{E}}}/\feps^{\textrm{\tiny{E}}}\,Z^{\textrm{\tiny{E}}})$ and $f_t=1/\feps^{\textrm{\tiny{E}}}$. We impose as integrability condition that the function $f_t$ has a fixed zero mode (which we set to unity with no loss of generality). As shown below this guarantees that the first variation of the full action vanishes for all variations preserving our boundary conditions. 

The variation of the boundary action \eqref{eq:cghs28} expands as
\begin{multline}
\delta I^{\textrm{\tiny{tw}}} = -\kappa\,\oint\extd t \,\big[\delta\big(f_tC\big) + C\,\delta f_t \\ 
\!\!- \langle (\partial_t x 
+ [G^{-1}\partial_t G,x]),G^{-1}\delta G \rangle + \partial_t\langle x, G^{-1}\delta G \rangle\big]
\label{eq:cghs29}
\end{multline}
with the bilinear Casimir~\footnote{%
Our BF theory has a linear Casimir, $Y$, and a bilinear Casimir, $C=X^+ X^- - YX$. Both are constant on-shell and invariant under all boundary condition preserving transformations, $\delta_\lambda Y = 0 = \delta_\lambda C$. See \cite{Ikeda:1993fh,Schaller:1994es} for more on BF-theories, its deformation to Poisson-sigma models and Casimirs appearing therein.}
\eq{ 
C=\tfrac12\,\langle B,\,B\rangle=\tfrac12\,\langle x,\,x\rangle\,. 
}{eq:cghs99} 
The second term in \eqref{eq:cghs29} vanishes on-shell since $C$ is constant and $f_t$ has a fixed zero mode, while the terms in the second line vanish on-shell or because they integrate to zero; thus, only the first term remains, which is integrable in field space, leading to the boundary action 
\eq{
I^{\textrm{\tiny{tw}}} = -\kappa\,\oint\extd t \,f_t\,C\,.
}{eq:cghs30}

For future purposes we define $f_t=\dot{f}$, where $f=t+\dots$ is a quasi-periodic function that has arbitrary Fourier modes but a fixed linear term. The full action
\eq{
\Gamma^{\textrm{\tiny{BF}}}[B,\,{\cal A}] = 
I^{\textrm{\tiny{BF}}}[B,\,{\cal A}] - \kappa\,\oint\extd t \,\dot{f}\,C
}{eq:cghs64}
has a well-defined variational principle, i.e., its first variation vanishes for all variations that preserve our boundary and on-shell conditions \eqref{eq:cghs4}-\eqref{eq:eom}.

Plugging the Euclidean version of the expression \eqref{eq:cghs27} for $x$ into the bilinear Casimir \eqref{eq:cghs99} yields
\eq{
C = -\frac{1}{(\dot{f})^2}\,\Big(\nlo  - \dot{f} \feta^{\textrm{\tiny{E}}} \lo +i\dot{f}\dot{\feta^{\textrm{\tiny{E}}}} + i\ddot{f}\feta^{\textrm{\tiny{E}}}\Big)
}{eq:cghs31}
where we used the relation $1/\feps^{\textrm{\tiny{E}}}=\dot{f}$. Defining additionally $\dot{g}:=i\feta^{\textrm{\tiny{E}}}\dot{f}$ the boundary action \eqref{eq:cghs30} with the expression for the Casimir \eqref{eq:cghs31} is nearly our final result. 
\eq{
I^{\textrm{\tiny{tw}}} = \kappa\,\oint\frac{\extd t}{\dot{f}}\,\big(\nlo  +i \dot{g}\lo + \ddot{g}\big)
}{eq:cghs33}
The boundary action \eqref{eq:cghs33} depends functionally on $f$ and $g$, both of which are boundary scalars, as evident from their transformation behavior under asymptotic symmetries, $\delta_\lambda{f}=\varepsilon\dot{f}$ and $\delta_\lambda{g}=\sigma+\varepsilon\dot{g}$. The latter also shows that $g$ is a phase under $U(1)$ gauge transformations.

As in the JT case \cite{Gonzalez:2018enk} we reparametrize the time coordinate along the boundary by a diffeomorphism $\tau:=f(t)$, where $\tau$ is our new (Euclidean) time coordinate with period $\beta$, and introduce the inverse of $f$ as a new field $h(\tau):=-f^{-1}(\tau)$. The other field, $g$, now also depends on $\tau$ and prime from now on means derivative with respect to $\tau$. Implementing this diffeomorphism in the boundary action \eqref{eq:cghs33} establishes the boundary action \eqref{eq:intro} announced in the introduction, where prime means $\tfrac{\extd}{\extd\tau}$.

The action \eqref{eq:intro} is our main result and constitutes the analogue of the Schwarzian. Since it has a geometric interpretation as group action for twisted warped coadjoint orbits, governed by the symmetries \eqref{eq:cghs40too}, we refer to it as ``twisted warped action''. This is analogous to the interpretation of the Schwarzian action as group action for Virasoro coadjoint orbits \cite{Witten:1987ty,Alekseev:1988ce}. We refer to \cite{Afshar:2019tvp} and refs.~therein for more on these mathematical aspects.

\newcommand{\newh}{{\tilde h}} 
\newcommand{\newg}{{\tilde g}} 

\paragraph{Solutions to twisted warped theory.} We study now classical solutions of the action \eqref{eq:intro} for constant representatives, $\nlo=\nlo_0$ and $\lo=\lo_0$.  The Hamiltonian formulation involves three canonical pairs ($i=1,2,3$)
\eq{
I^{\textrm{\tiny{tw}}}[q_i,\,p_i] = -\kappa\,\int\limits_0^\beta{\extd\tau}\,\big(p_i\,q_i^\prime - p_1\, p_2 - e^{q_1}\, p_3\big)\,.
}{eq:cghs57}
The relation to the original variables is  $q_3=\exp(i\lo_0\,h)$,  $q_2=g+ih\,\nlo_0/\lo_0$ while all other canonical variables are of auxiliary nature to get rid of higher derivatives. The interaction term with the exponential in $q_1$ also appears in the Schwarzian theory, see Eq.~(2.1) in \cite{Mertens:2017mtv}. The key difference is the kinetic term, $p_1^2$ for the Schwarzian and $p_1p_2$ for the twisted warped Hamiltonian.  

Solving the Hamiltonian equations of motion yields $q_3=h_0+h_1\,e^{i\tau/\tau_0}$ and $q_2=g_0-ig_1\,\tau+g_2\,e^{i\tau/\tau_0}$. These solutions depend on six integration constants, $g_0,g_1,g_2,h_0,h_1,\tau_0$, the latter playing the role of the periodicity, $\tau_0 = \frac{\beta}{2\pi}$. The integration constants $h_0$ and $g_0$ are constant shifts, while $h_1$ and $g_2$ are amplitudes in front of oscillating terms. The remaining constant, $g_1$, captures the non-periodicity of $q_2$ and is responsible for the on-shell action being non-zero, $I^{\textrm{\tiny{tw}}}[q_i,\,p_i]\big|_{\textrm{\tiny{EOM}}}=-2\pi\kappa\,g_1$.

\paragraph{Thermodynamics.} Assuming $g_1$ is independent from temperature allows to deduce the entropy $S=-I^{\textrm{\tiny{tw}}}[q_i,\,p_i]\big|_{\textrm{\tiny{EOM}}}=2\pi\kappa\,g_1$ from the on-shell action. Inserting all our definitions we recover the well-known fact that entropy is given by the dilaton at the horizon \cite{Gegenberg:1994pv}.
\eq{
S = 2\pi\kappa\,X\big|_{\textrm{horizon}}
}{eq:entropy}

The result for entropy \eqref{eq:entropy} can be derived along the lines of \cite{Grumiller:2015vaa,Grumiller:2017qao}. One aspect of this derivation is worth highlighting: the holonomy of $a$ along the thermal cycle must belong to the center of our gauge group for regularity (in order to have contractible thermal cycles). Assuming a single cover, we find that this regularity condition relates temperature $T=\beta^{-1}$ and charge
\eq{
\lo_0=2\pi T
}{holcond}
while mass $\nlo_0$ remains arbitrary. The label ``charge'' is justified for $\lo_0$ since the equations of motion imply $\lo_0=Y$ and $Y$ is the $U(1)$ charge. The label ``mass'' is justified for $\nlo_0$ as it is the subleading term in the metric \eqref{eq:cghs41} and since the associated function $\nlo$ transforms like a stress-tensor \eqref{eq:cghs40} in a twisted warped field theory \cite{Afshar:2015wjm}.

A peculiar aspect of $\cghs$ black hole thermodynamics is that the inverse specific heat (at fixed charge) vanishes, $C^{-1} = \frac{1}{T}\,\frac{\extd T}{\extd S}|_{\delta\lo_0=0} = 0$, since the Hawking--Unruh temperature $T$ trivially does not vary if the charge $\lo_0$ is kept fixed due to the relation \eqref{holcond}. This property is well-known \cite{Fiola:1994ir}, but will be crucial for the scaling limit from complex SYK. 

\paragraph{Scaling limit from complex SYK.} We turn now to the field theory side, starting with the complex SYK model \cite{Sachdev:2015efa,Maldacena:2016hyu,Davison:2016ngz,Chaturvedi:2018uov,Gu:2019jub}. The effective action governing the dynamics of the collective low temperature modes of complex SYK is given by (see \cite{Davison:2016ngz} and Eq.~(1.12) in \cite{Gu:2019jub})
\eq{
I^{\textrm{\tiny cSYK}} = \frac{NK}{2}\int\limits_0^\beta\extd\tau\big(g'+\tfrac{2\pi i{\cal E}}{\beta}h'\big)^2 - \frac{N\gamma}{4\pi^2}\int\limits_0^\beta\extd\tau\big\{\tan(\tfrac{\pi}{\beta}h);\tau\big\}
}{eq:cghs101}
where $\{f;\,\tau\}:=f'''/f'-\tfrac32(f''/f')^2$ is the Schwarzian derivative, $N$ is the (large) number of complex fermions, $NK$ is the zero temperature compressibility, $N\gamma$ is the specific heat at fixed charge and $\cal{E}$ is a spectral asymmetry parameter. The time-reparametrization field $h(\tau+\beta)=h(\tau)+\beta$ is quasi-periodic and the phase field $g(\tau)$, in the absence of winding, is periodic.

According to the thermodynamical discussion above we are interested in the limit $N\gamma\to\infty$ in order to obtain our action \eqref{eq:intro} as limit from the complex SYK effective action \eqref{eq:cghs101}. This is indeed possible by combining the actions \eqref{eq:intro} and \eqref{eq:cghs101} to the geometric action associated with the twisted warped Virasoro group, known as `warped Schwarzian' \cite{Afshar:2019tvp}, $I^{\textrm{wSch}}=I^{\textrm{\tiny{cSYK}}}+I^{\textrm{tw}}$.

Starting from the effective action \eqref{eq:cghs101} and shifting $g$ by \cite{Afshar:2019tvp} $g\to{g}-\frac{\kappa}{NK}({\log}h'+\frac{2\pi{i}}{\beta}h)$ yields the action $I^\textrm{wSch}$ with non-vanishing $\kappa$ and shifted specific heat parameter $\hat\gamma=\gamma+\frac{36\pi^2\kappa^2}{N^2K}$. Thus, our boundary action \eqref{eq:intro} emerges by sending both $\hat\gamma$ and $K$ to zero, while keeping fixed $\kappa$. At large $N$ this is achieved by the family of scaling limits
\eq{
\gamma = \gamma_0 N^{a} \quad  K = -K_0 N^{-b} \quad \kappa = \tfrac{N^{1+\frac{a-b}{2}}}{6\pi}\,\sqrt{\gamma_0 K_0}
}{eq:cghs102}
The constants $\gamma_0$ and $K_0$ are independent from $N$ and their product must be positive. The exponents $a>-1$, $b>1$ lead to infinite specific heat and vanishing zero temperature compressibility, respectively, in the large $N$ limit. Two simple choices are $a=b=2$, leading to $\kappa=\tfrac{N}{6\pi}\sqrt{\gamma_0K_0}$, and $a=0$, $b=2$, leading to  $\kappa=\tfrac{1}{6\pi}\sqrt{\gamma_0K_0}$. 

\paragraph{Conclusions.} We derived on the gravity side the boundary action \eqref{eq:intro} as a first step towards a two-dimensional model for flat space holography. We showed that the field theory side of our proposal for flat space holography emerges as a triple scaling limit of complex SYK: large $N$, large coupling (or small temperature) and large specific heat, while keeping fixed (with an adjustable scaling in $N$) the geometric mean of specific heat and zero temperature compressibility. As evident from \eqref{eq:cghs102}, this geometric mean (up to a factor $N^{1+\frac{a-b}{2}}/(6\pi)$) is the coupling constant $\kappa$ in \eqref{eq:intro}. 

Starting from our flat space holographic description numerous further research avenues can now be pursued, inspired by corresponding SYK-related results or by generic aspects of two-dimensional dilaton gravity (see \cite{Grumiller:2002nm} for a review and \cite{Grumiller:2006ja} for a list of models). Not intending to do justice to the vast literature on these subjects we highlight just one intriguing aspect, namely the role of chaos in flat space holography. By analogy to the AdS$_2$ case \cite{Jensen:2016pah,Gross:2019ach} we expect saturation of the chaos bound, i.e., a Lyapunov exponent given by $\lambda_L=2\pi\,T$. It should be rewarding to verify this through explicit calculations.

\acknowledgments

\paragraph{Acknowledgments.} We thank Arjun Bagchi, Bob McNees, Stefan Prohazka, Jakob Salzer and Carlos Valc\'arcel for collaborations on related subjects and for discussions. DG is particularly grateful to Jakob Salzer for joint unpublished work on BMS$_2$ in two-dimensional dilaton gravity.
This work was supported by the Austrian Science Fund (FWF), projects P~28751, P~30822 and P~32581. D.V. was supported in parts by the S\~ao Paulo Research Foundation (FAPESP), project 2016/03319-6, by the grant 303807/2016-4 of CNPq, by the RFBR project 18-02-00149-a and by the Tomsk State University Competitiveness Improvement Program. H.G. is funded by Fondecyt grant 11190427.

\begin{appendix}

\setcounter{equation}{0}
\renewcommand{\theequation}{S\arabic{equation}}

\section{Supplemental Material: Maxwell algebra}

\paragraph{Maxwell algebra and matrix representation.} The 1+1 dimensional Maxwell algebra is a central extension of the Poincar\'e algebra which has four generators associated to space (${\cal P}_1$) and time (${\cal P}_0$) translations, boosts ($\mathcal J$) and a central term ($\mathcal Z$);
\begin{align}
[{\mathcal P}_a,{\mathcal P}_b]=\epsilon_{ab}\,{\mathcal Z}  \qquad\qquad  [{\mathcal P}_a,{\mathcal J}]=\epsilon_a{}^b\,{\mathcal P}_b
\label{eq:maxwell}
\end{align}
where $\epsilon^{01}=1=-\epsilon_{01}$.  Changing the basis as $L_0={\mathcal J}$, $L_1={\mathcal P}_1-{\mathcal P}_0$, $J_{-1}={\mathcal P}_1+{\mathcal P}_0$ and $J_0=-2{\mathcal Z}$ yields
\begin{subequations}
\label{eq:warped}
\begin{align}
    [L_0,L_1]&=-L_1\\
    [L_0,J_{-1}]&=J_{-1}\\ 
    [L_1,J_{-1}]&=J_0
\end{align}
\end{subequations}
which coincides with a maximal subalgebra of the warped Witt algebra in the main text, re-displayed below. 
\begin{subequations}
\label{eq:cghs36}
\begin{align}
 [L_n,\,L_m]_{\textrm{\tiny Lie}} &= (n-m)\,L_{n+m}\\
 [L_n,\,J_m]_{\textrm{\tiny Lie}} &= -m\,J_{n+m}\\
 [J_n,\,J_m]_{\textrm{\tiny Lie}} &= 0
\end{align}
\end{subequations}
The relation to the light-cone generators in the main text is $L_1= P_+$, $J_{-1}= P_-$, $L_0=J$ and $J_0=Z$, in terms of which the algebra simplifies to
\eq{
[P_+,\,P_-] = Z \qquad\qquad [P_\pm,\,J] = \pm P_\pm\,.
}{eq:cghs6too}

Using light-cone generators a simple matrix representation for the algebra \eqref{eq:maxwell} in terms of $3\times 3$ matrices is 
\begin{align}
P_+ &=  
\begin{pmatrix}
0 & 1 & 0 \\
0 & 0 & 0 \\
0 & 0 & 0	
\end{pmatrix}  
& P_- &=  
\begin{pmatrix}
0 & 0 & 0 \\
0 & 0 & 1 \\
0 & 0 & 0	
\end{pmatrix} \displaybreak[1]\\
J &=  
\begin{pmatrix}
0 & 0 & 0 \\
0 & 1 & 0 \\
0 & 0 & 0
\end{pmatrix} 
& Z &=
\begin{pmatrix}
0 & 0 & 1 \\
0 & 0 & 0 \\
0 & 0 & 0	
\end{pmatrix} \,.\label{matrix_rep}
\end{align}
The bilinear form \eqref{eq:cghs3} cannot be represented by the simple matrix trace, since all traces of bilinear or quadratic expressions vanish, with the exception of $\tr J^2=1$. However, we recover the bilinear form \eqref{eq:cghs3} by introducing the adjoint matrices
\eq{
P_\pm^\dagger := P_\mp^T \qquad\quad J^\dagger := - Z^T \qquad\quad Z^\dagger := -J^T
}{eq:adjoint}
and then defining 
\eq{
\langle A,\,B\rangle := \tr\big(A^\dagger\,B\big) = \tr\big(A\,B^\dagger\big)\,.
}{eq:bilin_tr}
The adjoint \eqref{eq:adjoint} is involutive, i.e., $(A^\dagger)^\dagger=A$ for all generators $A=P_\pm,J,Z$, but it does not act in the usual way on products, $(AB)^\dagger\neq B^\dagger A^\dagger$ in general. 

\paragraph{Harmonic oscillator basis.} The 1+1 dimensional Maxwell algebra \eqref{eq:maxwell} [or \eqref{eq:warped} or \eqref{eq:cghs6too}] is identical to the harmonic oscillator algebra. This is seen explicitly by introducing the creation operator $a=L_1$, the annihilation operator $a^\dagger=J_{-1}$, the Hamiltonian $H=\frac1\hbar\,a^\dagger a=L_0$ and the central term $\hbar\,\unity=J_0$, with the usual commutation relations
\eq{
[a,\,a^\dagger] = \hbar\,\unity\qquad [H,\,a] = -a \qquad [H,\,a^\dagger] =  a^\dagger\,.
}{eq:oscillator}

\paragraph{Contraction from sl$(2)\oplus u(1)$.} The 1+1 dimensional Maxwell algebra \eqref{eq:cghs6too} can also be obtained as a contraction of sl$(2)\oplus u(1)$ with commutation relations
\eq{
[\hat L_+,\,\hat L_-]=2\hat L_0\quad[\hat L_\pm,\,\hat L_0]=\pm\hat L_\pm\quad[\hat J_0,\,\hat L_n] = 0
}{eq:sl2u1}
by first changing the basis
\eq{
\hat L_\pm = \frac1\eps\,P_\pm\qquad\quad \hat L_0 = J + \frac{1}{2\eps^2}\,Z\qquad\quad \hat J_0 = Z
}{eq:iw}
and then taking the limit $\eps\to 0$. 

Since sl$(2)\oplus u(1)$ is the gauge algebra in the gauge theoretic formulation of the charged JT model the existence of this contraction shows that a scaling limit from charged JT to $\cghs$ exists. This is the gravity version of the scaling limit of the complex SYK model studied in the main text.


\paragraph{Infinite lift and central extension.} The Maxwell algebra \eqref{eq:warped} has an infinite lift to the warped Witt algebra, see the Lie-bracket algebra  \eqref{eq:cghs36} of asymptotic Killing vectors. The warped Witt algebra has up to three non-trivial central extensions: the Virasoro central charge $c$, a twist term $\kappa$ and a $u(1)$-level $\hat K$. The centrally extended version of \eqref{eq:cghs36} reads
\begin{subequations}
\label{eq:lalapetz}
\begin{align}
    [L_n,\,L_m] &= (n-m)\,L_{n+m} + \frac{c}{12}\big(n^3-n\big)\delta_{n+m,0} \\
    [L_n,\,J_m] &= -m\,J_{n+m} - i\kappa\,\big(n^2-n\big)\,\delta_{n+m,\,0} \\
    [J_n,\,J_m] &= \frac{\hat K}2\,n\,\delta_{n+m,\,0}\,.
\end{align}
\end{subequations}
Note that the Maxwell algebra \eqref{eq:warped} is a subalgebra of \eqref{eq:lalapetz} that is blind to all three central extensions.

It is customary to give the algebra \eqref{eq:lalapetz} different names depending on which of the central extensions is non-zero. If $\kappa\neq 0$ we call the algebra ``twisted warped conformal'', regardless of the values of $c$ or $\hat K$; if $c\neq 0, \kappa\neq 0, \hat K\neq 0$ we call the algebra ``twisted warped Virasoro''; if $c\neq 0, \kappa=0, \hat K\neq 0$ we call the algebra ``warped Virasoro'';  if $c=0, \kappa\neq 0, \hat K=0$ we call the algebra ``twisted warped Witt''; if $c=\kappa=\hat K=0$, as in \eqref{eq:cghs36}, we call the algebra ``warped Witt''. 

The twisted warped Virasoro algebra can be mapped to the warped Virasoro algebra (with central charge $c\to c-\frac{24\kappa^2}{\hat K}$) by a change of basis, namely by first twisting the Virasoro generators, $L_n\to L_n + i\,\frac{2\kappa}{\hat K}\,n\,J_n$, and then shifting both zero modes $L_0\to L_0+\Delta_0$, $J_0\to J_0+Q_0$ with suitably chosen constants $\Delta_0$ and $Q_0$. 
The twist of the Virasoro generators no longer works if the $u(1)$ level vanishes, $\hat K=0$, so there is no regular way to eliminate the twist term $\kappa$ from the twisted warped Witt algebra. 

\paragraph{Singular limit.} A singular limit maps the warped Virasoro algebra (after twisting with some $\kappa$, i.e., inverting the map above from warped Virasoro to twisted warped Virasoro) to the twisted warped Witt algebra, namely $\hat K\to 0$, $c\to\infty$ while keeping fixed the geometric mean of central charge and $u(1)$-level,
\eq{
\kappa = \sqrt{-\frac{c\hat K}{24}}
}{eq:gm}
This is the algebraic version of the scaling limit we performed in \eqref{eq:cghs102}, with central charge $c=\frac{N\gamma}{3\pi^2}$ playing the role of specific heat, $u(1)$-level $\hat K=2NK$ playing the role of zero temperature compressibility and the twist term $\kappa$ is identical to the coupling constant $\kappa$ in the main text.

The twisted warped Witt algebra case, $c=0$, $\kappa\neq 0$, $\hat K=0$, is the one associated with our main result, the boundary action \eqref{eq:intro} with the symmetries \eqref{eq:cghs40too}. In a holographic context the twisted warped Witt algebra was discussed first in \cite{Afshar:2015wjm}, including a derivation of a Cardy-like entropy formula.
 \enlargethispage{0.5truecm}

\end{appendix}

\bibliographystyle{fullsort}
\bibliography{review2}

\end{document}